\documentclass[prd,preprint,linenumbers,amssymb,amsmath,aps,nofootinbib,superscriptaddress]{revtex4}

\usepackage[utf8]{inputenc} 
\usepackage[T1]{fontenc}    
\usepackage{booktabs}       
\usepackage{amsfonts}       
\usepackage{nicefrac}       
\usepackage{microtype}      
\usepackage{graphicx}
\usepackage{natbib}
\usepackage{doi}
\usepackage{mathrsfs}
\usepackage{caption}
\usepackage{overpic}

\begin{document}
\title{New constraints on singlet scalar dark matter model with LZ, invisible Higgs decay and gamma-ray line observations}

\author{Yang Yu}
\affiliation{Key Laboratory of Dark Matter and Space Astronomy, Purple Mountain Observatory, Chinese Academy of Sciences, Nanjing 210023, China}
\affiliation{School of Astronomy and Space Science, University of Science and Technology of China, Hefei 230026, China}

\author{Tian-Peng Tang}
\email{tangtp@pmo.ac.cn}
\affiliation{Key Laboratory of Dark Matter and Space Astronomy, Purple Mountain Observatory, Chinese Academy of Sciences, Nanjing 210023, China}
\author{Lei~Feng}
\email{fenglei@pmo.ac.cn}
\affiliation{Key Laboratory of Dark Matter and Space Astronomy, Purple Mountain Observatory, Chinese Academy of Sciences, Nanjing 210023, China}
\affiliation{School of Astronomy and Space Science, University of Science and Technology of China, Hefei 230026, China}
\affiliation{Joint Center for Particle, Nuclear Physics and Cosmology,  Nanjing University -- Purple Mountain Observatory,  Nanjing  210093, China}

\begin{abstract}
The singlet scalar dark matter (DM) model is a minimal extension of the Standard Model (SM).
This model features only two free parameters: the singlet scalar mass $m_S$ and the quartic coupling $a_2$ between the singlet scalar and the SM Higgs doublet.
Previous studies have suggested that only the resonant region remains viable under combined constraints.
In this work, we revisit and refine the constraints using data from the direct detection experiment LUX-ZEPLIN(LZ), invisible Higgs decay measurements, and gamma-ray line observations conducted by Fermi-LAT and DAMPE.
Our findings indicate that the latest results from LZ have significantly strengthened the constraints on the model parameters.
We demonstrate that only a narrow parameter region remains viable, specifically $60.5\,\text{GeV}< m_S < 62.5\,\text{GeV}$ and $1.7\times10^{-4}<a_2< 4.7\times10^{-4}$.
This suggests that a three-fold improvement in the current LZ results would thoroughly examine the parameter space below $1\,\rm{TeV}$ for the singlet scalar DM model.  

\end{abstract}

\maketitle

\section{Introduction}
Many cosmological and astronomical observations provide compelling evidence for the existence of dark matter (DM), yet its nature remains a mystery.
The Weakly Interacting Massive Particle (WIMP) is considered one of the most promising candidates for DM due to its ability to accurately reproduce the observed DM relic abundance through the so-called thermal relic mechanism~\cite{Bergstrom:2000pn, Bertone:2004pz, Arcadi:2017kky}.
In recent years, indirect detection of DM through cosmic-ray searches has revealed several intriguing excesses, including the Galactic Center GeV gamma-ray excess (GCE)~\cite{Hooper:2010mq}, AMS-02 antiproton excess~\cite{Cui:2016ppb, Cuoco:2016eej}, and electron/positron excess~\cite{Chang:2008aa,HESS:2009chc,Fermi-LAT:2009yfs,PAMELA:2008gwm,Fermi-LAT:2011baq,AMS:2013fma,AMS:2014xys}.
These anomalies can be coherently interpreted as resulting from WIMP annihilation into SM final states\cite{Feng:2013zca,Zhou:2014lva, Calore:2014xka, Daylan:2014rsa,Ge:2017tkd, Cui:2018klo, Cholis:2019ejx,Ge:2020tdh,Abdughani:2021oit,Fan:2022dck,Chen:2024njd, Fan:2024wvo,Fan:2024rcr}.

The simplest model for a WIMP candidate is the Singlet Scalar Higgs portal model, which only introduces a new massive real scalar field $S$ that does not carry any charges under the Standard Model (SM) gauge groups~\cite{Silveira:1985rk,McDonald:1993ex,Burgess:2000yq,Davoudiasl:2004be,OConnell:2006rsp}.
The discrete $Z_2$ symmetry is introduced to ensure the stability of the new scalar field.
Under this symmetry, the SM Higgs doublet $H$ is $Z_2$-even while the new scalar $S$ is $Z_2$-odd.
The symmetries of the Standard Model restrict the renormalizable interactions between $S$ and the SM fields to terms of the form $S^2H^\dagger H$ in the Lagrangian, as indicated by Eq.(\ref{eq_1}).
This interaction allows $S$ to couple with the SM via the Higss field, resulting in various observable effects.
Specifically, $S$ can annihilate into SM particles via the Higgs portal.
This annihilation process is essential for the DM thermalization in the early Universe and can produce annihilation signals in the Milky Way today.
Furthermore, DM-quark or gluon elastic scattering mediated by Higgs exchange could yield detectable signals in direct detection experiments.
Finally, the SM Higgs boson has the potential to decay into a pair of $S$, which could be tested in collider experiments.
Early related studies can be found in Refs.~\cite{Profumo:2010kp,1306.4710,Feng:2014vea,1407.6588,1609.03551,1705.07931,2011.13225,Ren:2014mta,2101.02507,2305.11937,2406.01705}.

Due to its simplicity, there are only two free model parameters, making it easier to constrain.
By integrating various detection methods, we can achieve a more comprehensive validation and constraint of the singlet scalar DM model, which is the primary objective of this work.
In light of the recent direct detection data from the LZ collaboration~\cite{LZCollaboration:2024lux}, the indirect detection results from gamma-ray line spectrum induced by DM annihilation from Fermi-LAT~\cite{Fermi-LAT:2015kyq} and DAMPE~\cite{DAMPE:2021hsz}, as well as the latest measurements of the Higgs invisible decay width from the ATLAS collaboration~\cite{ATLAS:2022yvh}, we identify parameter regions with the following intriguing features:
(i) Only a narrow parameter region remains viable, i.e. $60.5\,\text{GeV} < m_S < 62.5\,\text{GeV}$ and $1.7\times10^{-4}<a_2< 4.7\times10^{-4}$. 
(ii) A three-fold improvement over the current LZ results would effectively test this parameter space. Future direct detection experiments, such as DARWIN~\cite{DARWIN:2016hyl}, have the potential to achieve this level of precision.

The remaining sections of the paper are organized as follows. In Sec. \ref{model}, we provide a brief introduction to the singlet scalar DM model. Next, we present our main results in Sec. \ref{Result}. Finally, in Sec. \ref{conclusion}, we summarize our findings and discuss their implications.

\section{Model Setup}
\label{model}
The Lagrangian of the singlet scalar DM model can be written as follows~\cite{Profumo:2010kp, Profumo:2007wc}:

\begin{equation}
    \mathcal{L}=\mathcal{L}_\text{SM}+\frac{1}{2} \partial_{\mu }S\partial^{\mu }S-\frac{b_{2}}{2}S^{2} -\frac{b_{4}}{4}S^{4}-a_{2}S^{2}H^{\dagger}H
    \label{eq_1}.
\end{equation}
Here, $\mathcal{L}_\text{SM}$ represents the SM Lagrangian. $S$ denotes the real singlet scalar field, while $H$ is the SM Higgs doublet field.
The parameters $b_{2}, b_{4}$ and $a_{2}$ correspond to the mass of $S$, the self-coupling of $S$, and the coupling between $S$ and $H$, respectively.
Due to the introduction of the $Z_2$ symmetry, $S$ is stabilized and never obtains a vacuum expectation value (VEV), i.e., $\langle S\rangle=0$~\cite{Profumo:2007wc}.
After spontaneous symmetry breaking, we can express $H^{\dagger}=1/\sqrt{2}(h+v,0)$, where $h$ is the physical Higgs boson and $v\approx 246\,\rm{GeV}$ is the VEV of the SM Higgs field. The scalar potential is given by
\begin{equation}
    V(h,S)=-\frac{\mu ^{4}}{4\lambda } -\mu ^{2}h^{2}+\lambda v h^{3}+\frac{\lambda}{4} h^{4}+\frac{1}{2}(b_{2}+a_{2}v^{2})S^{2}+\frac{b_{4}}{4}S^{4}+a_{2}v S^{2}h+\frac{a_{2}}{2}S^{2}h^{2},
\end{equation}
where $\mu^{2}<0$, $\lambda$ is the quartic coupling of Higgs, and $(-\mu^{2}/\lambda)^{1/2}=v$.
This potential is bounded from below at tree level, given that $\lambda, b_{4} \geq 0$ and $\lambda b_{4} \geq a_{2}^{2}$ for negative $a_{2}$.
The mass of $S$ can expressed as
\begin{equation}
   m_{S}^{2}=b_{2}+a_{2}v^{2}.
\end{equation}
The fourth-order coupling term $b_{4}$ does not affect the observable physical phenomena \cite{Profumo:2010kp}.
Instead, the observable characteristics of this model depend solely on the two free parameters $a_{2}$ and $m_{S}$.

\begin{figure}[htbp]
     \centering
     \includegraphics[width=0.3\linewidth]{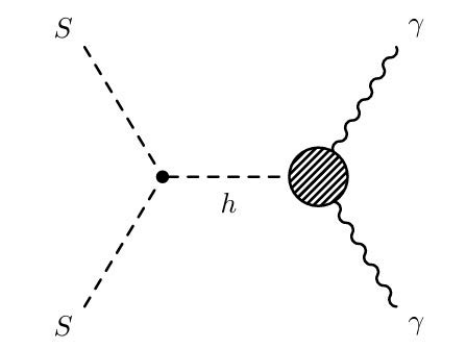}
     \caption{The Feynman diagram illustrating the annihilation of two singlet scalar DM into two photons through the Higgs portal.}
     \label{fem}

\end{figure}

For direct detection, we utilized the latest data from the LZ experiment to constrain the model parameters.
The DM-nucleus spin-independent cross section for this model is given by \cite{Cheung:2012xb}
\begin{equation}
   \sigma _\text{SI}=\frac{a_{2}^{2}m_{N}^{4}f^{2}}{\pi m_{S}^{2}m_{h}^{4}}  ,
\end{equation}
where $m_{N}$ is the nucleon mass and $f$ is the form factor, which we set to $1/3$ \cite{Farina:2009ez,Giedt:2009mr}.

The indirect detection constraints arise from the gamma-ray line spectrum. We focus on the DM annihilating into two photons, represented by the Feynman diagram in Fig.\ref{fem}. The cross section for this process is given by
\cite{1306.4710,LHCHiggsCrossSectionWorkingGroup:2011wcg,Gunion:1989we,Djouadi:2005gi}
\begin{equation}
   \langle\sigma v\rangle_{\gamma \gamma}=a_{2}^{2} \frac{\alpha^{2}}{32 \pi^{3}} \frac{s}{\left(s-m_{h}^{2}\right)^{2}+\Gamma^{2} m_{h}^{2}}\left|\sum_{i} N_{c i} e_{i}^{2} F_{i}\right|^{2},
   \label{eq_5}
\end{equation}
where $\alpha$ is the electromagnetic fine structure constant, $s$ is the square of the center-of-mass energy, and $\Gamma$ is the decay width of the Higgs boson.
The index $i$ denotes either a fermion or a gauge boson, $N_{ci}$ is its color multiplicity, and $e_{i}$ is its electric charge.

Besides, we set the branching ratio of the Higgs invisible to be 0.145 \cite{ATLAS:2022yvh}. The corresponding coupling coefficient is expressed as \cite{Feng:2014vea}
\begin{equation}
   a_{2}^2 = \frac{8 \pi m_h R_\text{inv} \Gamma_\text{vis}}{(1 - R_\text{inv}) v^2} \cdot \sqrt{\frac{m_h^2}{m_h^2 - 4m_S^2}},
\end{equation}
where $\Gamma_\text{vis} = 4.07\,\text{MeV}$ is the visible decay width of the Higgs boson.

\section{Result}
\label{Result}

\begin{figure}[htbp]
    \centering
    \includegraphics[width=0.9\linewidth]{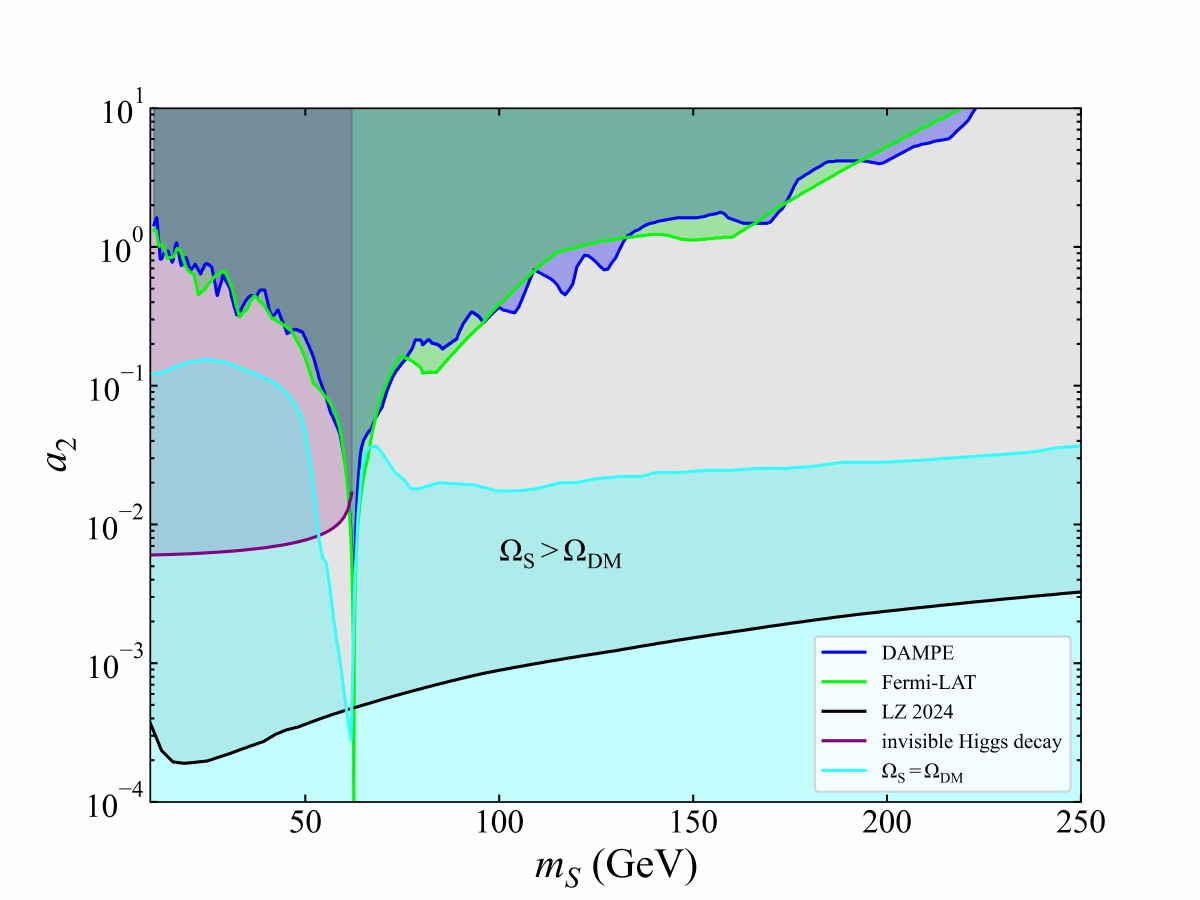}
    \label{fig1}
    \caption{Combined constraints on model parameters ($m_S,a_2$) from direct detection experiment LZ (black line)~\cite{LZCollaboration:2024lux}, gamma-ray spectra from Fermi-LAT (green line)~\cite{Fermi-LAT:2015kyq} and DAMPE (blue line)~\cite{DAMPE:2021hsz}, and invisible Higgs decay measurements (purple line)~\cite{ATLAS:2022yvh}.
    The regions shaded in cyan indicate areas where $\Omega_S>\Omega_\text{DM}$.
    Only a narrow region near the resonant area $m_{S} \simeq m_{h}/2$ remains viable.}
    \label{fig1}
\end{figure}

Some previous studies have already constrained the parameter space of this model  \cite{Feng:2014vea, Profumo:2010kp}. However, several new results in DM detection have emerged, including the latest findings from the direct detection experiment LZ \cite{LZCollaboration:2024lux}, measurements of invisible Higgs decay \cite{ATLAS:2022yvh}, and gamma-ray line searches conducted by Fermi-LAT \cite{Fermi-LAT:2015kyq} and DAMPE \cite{DAMPE:2021hsz}. 
In this study, we revisit and refine the constraints on the singlet scalar DM model using these new DM search results.

In Fig.~\ref{fig1}, we present the results for the DM mass $m_S$ versus coupling coefficient $a_2$.
The black line indicates the constraints obtained from the LZ experiment, while the cyan line represents the parameters that yield the correct DM relic density, derived from the numerical solution of the Boltzmann equation. Here, we refer to the parameters corresponding to the correct DM relic density in Ref. \cite{Feng:2014vea,1306.4710}.
As noted in our previous work \cite{Feng:2014vea}, advancements in the sensitivity of direct detection experiments have effectively excluded the region where $m_{S}$ > $110\,\text{GeV}$.
On the other hand, we utilize the new gamma-ray line searches from Fermi-LAT~\cite{Fermi-LAT:2015kyq} and DAMPE \cite{DAMPE:2021hsz}, concentrating on the $\gamma \gamma$ channel in NFW profile.


\begin{figure}[htbp]
    \centering
    \includegraphics[width=\linewidth]{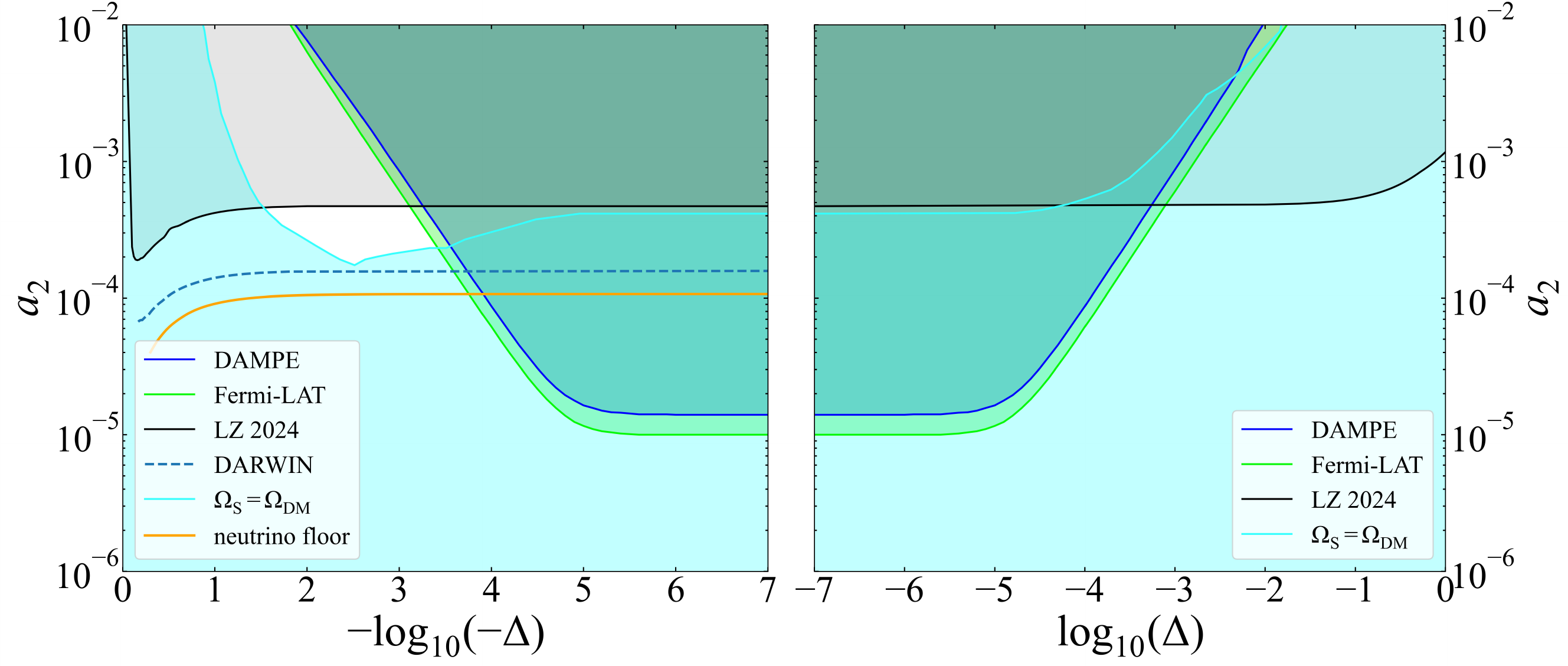}
    \caption{Enlarging the resonant region through coordinate transformation $ \Delta =(2m_{S}-m_{h})/m_{h} $. The blue dashed line indicates the projected sensitivities of DARWIN~\cite{DARWIN:2016hyl},  while the orange line represents the neutrino floor\cite{OHare:2021utq}. The definitions of the other colors are consistent with those in Fig.~\ref{fig1}.}   
    \label{fig2}
\end{figure}

In the Higgs resonance region, i.e., $m_{S} \simeq m_{h}/2$, the allowed range of the coupling coefficient $a_2$ that satisfies the correct DM relic density exhibits a sharp drop.
As show in Fig.~\ref{fem}, in this model, DM annihilation occurs via the $s$-channel Higss exchange. According to Eq.~(\ref{eq_5}), when $s = (2m_S)^2 \simeq m_h^2$, the denominator term reaches its minimum, significantly enhancing the DM annihilation cross-section. 
Consequently, this resonance mechanism enables even a small coupling ($a_2 \sim 10^{-4}$) to drive sufficient annihilation in the early universe, ensuring consistency with the relic density constraint.

Notably, the Higss resonance annihilation can generate a distinctive gamma-ray line signal, to which DM indirect detection experiments are highly sensitive. To better illustrate the latest constraints from DAMPE and Fermi-LAT gamma-ray line searches on the resonance parameters, we define $\Delta = (2m_{S} - m_{h})/m_{h}$ following Ref.~\cite{Feng:2014vea} to focus on the resonant region, with the corresponding results presented in Fig.~\ref{fig2}.
The gamma-ray spectrum data strongly exclude the parameter space for masses slightly above $m_h/2$, but the exclusion power is weaker for the region just below $m_h/2$.
This is because, during the freeze-out stage in the early universe, DM $S$ carries non-negligible kinetic energy, allowing the resonance condition $s = m_h^2$ to be met even for $m_S < m_h/2$.
As a result, the relic density constraint favors $m_S$ slightly below $m_h/2$. However, in the present-day Galactic center, where DM is nearly non-relativistic, the resonance condition holds at $m_S \approx m_h/2$, making indirect detection constraints more effective for masses above the resonance.


For DM masses slightly below $m_h/2$, the direct detection experiments LZ~\cite{LZCollaboration:2024lux} is particularly effective in excluding a substantial portion of the parameter space. The viable region is defined by the range $-10^{-3.4}< \Delta < -10^{-1.5}$, corresponding to $60.5\, \text{GeV} < m_{S} < 62.5\, \text{GeV}$.
Within this region, the coupling parameter is approximately $1.7\times10^{-4}<a_2< 4.7\times10^{-4}$.
If the LZ data were to improve by a factor of three, it could effectively access these parameter spaces. Moreover, if the DARWIN~\cite{DARWIN:2016hyl} experiment achieves its expected precision, it could thoroughly probe this region.

\section{Discussion and Conclusions}
\label{conclusion}

The singlet scalar model is one of the simplest extensions to the SM, featuring only two free parameters.
In this work, we examine this model with the current DM experiments and find that DM direct detection can significantly enhance the constraints on the parameter space just below half the mass of the Higgs boson.
Our analysis reveals that only a narrow viable region remains, specifically $60.5\, \text{GeV} < m_{S} < 62.5\, \text{GeV}$  and $1.7\times10^{-4}<a_2< 4.7\times10^{-4}$. 
An improvement by a factor of three in the current LZ results would effectively close in on this parameter space.
Future direct experiments, such as DARWIN, are expected to achieve this level of precision.
Then, this model will undergo its final judgment.



\section*{Acknowledgments}
This work is supported by the National Key R\&D Program of China (Grants No. 2022YFF0503304), the National Natural Science Foundation of China (12373002, 12220101003, 11773075), the Youth Innovation Promotion Association of Chinese Academy of Sciences (Grant No. 2016288), and the Jiangsu Province Post Doctoral Foundation (No.2024ZB713).

\end{document}